\documentclass[a4paper]{jpconf}
\usepackage{graphicx}
\usepackage{xcolor}
\begin{document}
\bibliographystyle{iopart-num}
\title{Joint machine learning analysis of muon spectroscopy data from different materials}

\author{T Tula$^1$, G M\"oller$^1$, J Quintanilla$^1$, S R Giblin$^2$, A D Hillier$^3$, E E McCabe$^{1,4}$, S Ramos$^1$, D S Barker$^{1,5}$ and S Gibson$^1$ }
\address{$^1$ School of Physical Sciences, University of Kent,
Park Wood Rd, Canterbury CT2~7NH, United Kingdom}
\address{$^2$ School of Physics and Astronomy, Cardiff University, Cardiff CF24 3AA, United Kingdom}
\address{$^3$ 
ISIS Facility, STFC Rutherford Appleton Laboratory, Chilton, Didcot Oxon, OX11 0QX, United Kingdom}
\address{$^4$ Department of Physics, Durham University, Durham, DH1 3LE, United Kingdom}
\address{$^5$ School of Physics and Astronomy, University of Leeds, Leeds, LS2 9JT, United Kingdom}

\ead{tt343@kent.ac.uk}

\newcommand{\comm}[1]{}
\newcommand{\commred}[1]{{\color{red}#1}}

\begin{abstract}
Machine learning (ML) methods have proved to be a very successful tool in physical sciences, especially when applied to experimental data analysis. Artificial intelligence is particularly good at recognizing patterns in high dimensional data, where it usually outperforms humans. Here we applied a simple ML tool called principal component analysis (PCA) to study data from muon spectroscopy. The measured quantity from this experiment is an asymmetry function, which holds the information about the average intrinsic magnetic field of the sample. A change in the asymmetry function might indicate a phase transition; however, these changes can be very subtle, and existing methods of analyzing the data require knowledge about the specific physics of the material. PCA is an unsupervised ML tool, which means that no assumption about the input data is required, yet we found that it still can be successfully applied to asymmetry curves, and the indications of phase transitions can be recovered. The method was applied to a range of magnetic materials with different underlying physics. We discovered that performing PCA on all those materials simultaneously can have a positive effect on the clarity of phase transition indicators and can also improve the detection of the most important variations of asymmetry functions. For this joint PCA we introduce a simple way to track the contributions from different materials for a more meaningful analysis.
\end{abstract}

\section{Introduction}
In recent years, machine learning tools have become increasingly familiar to the scientific community, but they are not yet regarded as standard practice for analysing small-scale experimental measurements in condensed matter physics. The strength of ML techniques lies in their ability to recognise patterns and correlations in data \cite{Zdeborova:2017eo, Carleo:2019hc, geron2019}, which is often the goal of scientific enquiries (such as the search for phase transitions). Usually, for the ML methods to work, a large data set is required and since data acquisition can be costly for many experiments (in both time and resources), it is understandable that ML tools are not always thought to be a suitable tool. In this work, we want to test whether experimental muon spectroscopy measurements can be analysed with ML even when using smaller sets of data. Specifically we apply a linear principal component analysis algorithm to muon spectroscopy measurements \cite{intro-muons} in order to find signs of phase transitions for a range of different materials.

In this article we report on a new development from our previous research \cite{tula2021machine}, in which we demonstrated that PCA can be successfully applied to a range of different materials simultaneously, thus increasing the effectiveness of this ML method, even when the underlying physics of the materials are vastly different. 
%We now learned that in order to properly analyse the data in that case, we need to keep track of certain weights in PCA algorithm, that tells us how important the output of different PCs are for each material. More on that will be explained in ``results'' section.
To have more insight into the PCA results in this case, we present a cumulative principal component score metric, which tells us how important each PC output is for a given material. That allows us to look at phase transition indicators in correct principal component scores, and this improved our previous analysis of magnetic materials.

\section{Methods}
\subsection{Muon spectroscopy experiments}
The main quantity to analyse from muon spectroscopy experiments is the asymmetry function, which tells us the distribution of spin directions of muons implanted into the sample. This can be directly linked to the internal magnetic field of the material studied \cite{Schenck1985Jan, s_f_j_cox_implanted_1987, Blundell1999, Lee1999}. We are interested in studying changes in the asymmetry functions at different temperatures, which can indicate phase transitions. A more comprehensive introduction to muon spectroscopy can be found in \cite{intro-muons, tula2021machine}.
\subsection{PCA}
Principal component analysis is an unsupervised machine learning technique \cite{geron2019, jolliffe2016principal}, which means that we are not required to label the input data, and given that there are some correlations between asymmetry functions, PCA will be able to find those. Applying PCA is equal to performing singular value decomposition on a matrix containing all asymmetry functions $\mathbf{A}$. Principal components (PC) are given by left-singular vectors of $\mathbf{A}$, with singular values signifying the importance of each PC. By looking at projections of asymmetry functions into the most important PCs, we can see changes in the behaviour that could indicate phase transitions. More on the PCA method as applied to $\mu$SR data can be found in our previous article \cite{tula2021machine}
%We can represent each asymmetry function $A_j(t)$ as a point $A_j$ in a multi-dimensional space (with dimensions represented by function values at each time window: $t_1$, $t_2$, ...). We expect that the data will be correlated, which means that the points $A_j$ will be lying close to some low-dimensional subspace ($\pi_1$ and $\pi_2$ in figure \ref{fig:1} c)$\,$). PCA performs a rotation on the initial coordinate space to a new orthonormal basis in which only few directions are required to capture most of the covariance of the data, so it allows us to find that low dimensional subspace and provide us with useful basis. The vectors defining this new basis are called principal components (PCs) and the projections of measurements onto PCs are known as PC scores. By looking at the most important principal components (the ones that hold most of the covariance) and their scores dependence on temperature, we can find even subtle changes in asymmetry scores shapes. 
\comm{
\begin{figure}[b]
    \centering
    \includegraphics[width=1\linewidth]{/figure2a_2f.eps}
    \caption{Example of PCA performed on muoon spectroscopy data: a)-c) Three asymmetry curves from the same sample of BaFe$_2$Se$_2$O at different temperatures. The red curve correspond to the reconstruction given by equation (\ref{eq:2}); d) Average of all measurements (equation (\ref{eq:1})); e) Shapes of the two most important principal components; f) graph of 1st PC score vs 2nd PC score showing the position of curves a)-c). Figure taken from \cite{tula2021machine}}
    \label{fig:2}
\end{figure}

We present an example of PCA applied to muon spectroscopy functions in figure \ref{fig:2}. From 23 asymmetry curves, there are 3 examples plotted. For the method to work we need to first to subtract the average of the functions defined as
\begin{equation} \label{eq:1}
    AV[\{A_j(t_i)\}_j](t_i) = \frac{1}{N}\sum_{j=1}^{N} A_j(t_i),
\end{equation}
where $N$ is a number of measurements. After performing PCA we find that the two first PCs hold most of the covariance of the data. From figure \ref{fig:2} e) we can see that the 1st PC looks almost like a linear function, while the 2nd PC is curved. Because PC scores are projections of asymmetry functions onto PCs we can write
\begin{equation} \label{eq:2}
    A_j(t) \approx AV(t) + \mbox{PC}_1\mbox{score}(j) \times \mbox{PC}_1(t) + \mbox{PC}_2 \mbox{score}(j) \times \mbox{PC}_2(t).
\end{equation}

The contributions from the rest of PCs can be neglected. With that it is evident that because of the curvature of the asymmetry function b), it will have a larger 2nd PC score than either a) or c). Additionally, the low temperature asymmetry function a) does have a different slope than the high temperature one, which is reflected in their different 1st PC score. These results can be seen on the 1st PC score vs 2nd PC score graph (figure \ref{fig:2} f)$\,$), where we can distinguish three different groups of points, which correspond to different phases of the material.
}
\section{Results and Discussion}
We applied PCA to a range of different materials. For each of them we had around 10-30 asymmetry functions, measured at different temperatures and at vanishing external magnetic field throughout. Among the systems studied are time-reversal symmetry breaking superconductors (LaNiGa$_2$, LaNiC$_2$, LaNi$_{0.9}$Cu$_{0.1}$C$_2$, LaNi$_{0.75}$Cu$_{0.25}$C$_2$), a proposed spin liquid (LuCuGaO$_4$) and a layered antiferromagnet (BaFe$_2$Se$_2$O). We initially tried performing the analysis for each material individually, but found out that for some materials the small number of measurements resulted in most PCs having a significant contribution to covariance. To overcome that, we performed PCA on the joint dataset of measurements from all materials simultaneously. This improved the visibility of phase transitions for those measurements that were previously hard to analyse, and did not significantly alter the other results. This is remarkable given that the materials in the data set have very different underlying physics. We can interpret this result as follows: by feeding all the data to the PCA simultaneously, the algorithm is not learning anything new about any particular material, but it is learning more about the experimental technique being employed in all cases ($\mu$SR). A comparison of the shapes of the ensuing principal components -- with PCA performed on one material at a time or simultaneously on all of them -- can be found in our previous work \cite{tula2021machine}.

We now demonstrate that more insight can be gained from the results of a PCA analysis performed on multiple samples by calculating the total score of a given PC for each individual material. Unlike analysis for a single material, where PC scores are naturally ordered in terms of decreasing significance, in a sample of several materials considered jointly, the order of significance can depend on the material: Some PC shapes might be dominant in most of materials, meaning that they will have large contribution to the covariance, but for other materials another PC might be more prominent. To interpret the analysis, we should always examine the most important PC(s) for each sample. To identify the relevant scores, we define a cumulative principal component score (cPCS) as
\begin{equation} \label{eq:3}
    \mbox{cPCS}(i; n) = \sum_{j=1}^{N_{i}}|\mbox{PCS}_{n}^{i}(j)|,
\end{equation}
where $i$ iterates over different materials, $n$ stands for $n$-th principal component, PCS$_n^{i}(j)$ corresponds to $n$-th principal component score for a $j$-th measurement of a given material and $N_{i}$ stands for number of measurements for $i$-th material. We then use this measure to identify the principal component scores with the largest cPCS for each material.

\begin{figure}[t]
    \centering
    \includegraphics[width=1\linewidth]{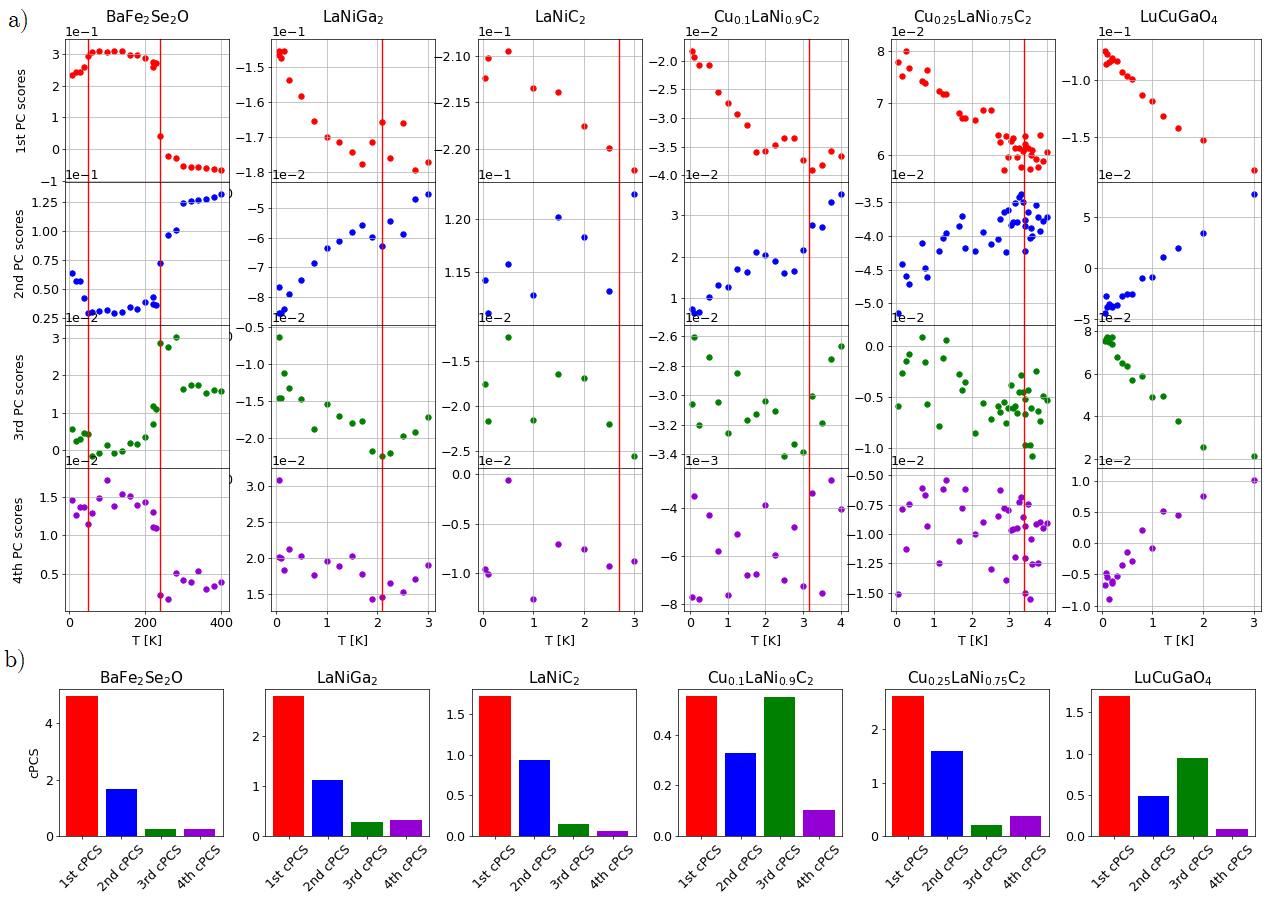}
    \caption{a) Temperature dependence for principal component scores when applying PCA to multiple materials simultaneously. Red vertical lines correspond to expected phase transitions, taken from the literature; b) Cumulative principal component scores for all materials.}
    \label{fig:3}
\end{figure}
Our findings are presented in figure \ref{fig:3}, which is an extension of figure 5 from \cite{tula2021machine}. We present the temperature dependence of the first four principal component scores for all the materials in a), and their cPCS in b). From the cPCS we can see that although the 1st PC has the largest contribution to all of measurements, the 2nd PC is less important than the 3rd one for Cu$_{0.1}$LaNi$_{0.9}$C$_2$ and LuCuGaO$_4$. If we compare the principal component scores for Cu$_{0.1}$LaNi$_{0.9}$C$_2$, we can clearly see that the 3rd PC has a much more visible indication of the expected phase transition (red vertical line) than the 2nd or even 1st one. 
For LuCuGaO$_4$ material, we expected improvement from our previous joint analysis, and although we find that 3rd PC is more important than 2nd, there seems to be no feature indication at 0.6 K, which we have found in principal component analysis of just this material (\cite{tula2021machine}, figure 4, last column, third row). This tells us that joint analysis is most helpful whenever the separate PCA is not successful, which we can always tell by looking at scree plots. In the case of LuCuGaO$_4$, the cPCS metric looks worse than scree plot of its separate analysis (\cite{tula2021machine}, figure 4, last column, last row), and we should take that into consideration when analysing the material. One important note here is that adding LuCuGaO$_4$ did not affect negatively the results for other materials, which we have showed previously \cite{tula2021machine}.
%In LuCuGaO$_4$ we find a similar behaviour for the 1st and 3rd PC score, although for the latter there is slight kink in the function above 2 K.
\section{Conclusions}
We have suggested a new analysis method to extend our recently propsed PCA of $\mu$SR data \cite{tula2021machine}. We are more confident that performing a joint PCA of measurements taken for different materials is a viable method of analysis, given that for each set of measurements we are looking at the right PC scores, which can be ensured by calculating a simple metric -- cumulative principal component scores. We should also compare the cPCS to scree plots from  separate analyses conducted for each material, in case the simultaneous PCA might not improve the clarity of each individual study. 
%-- although adding such materials for joint analysis does not seem to negatively affect our results for the rest of the materials. 
The ability to thus enhance our understanding gained from a new experiment based on existing data goes beyond the possibilities of preexisting approaches, where data for each material is necessarily analyzed and fitted in isolation, and overarching commonalities are anticipated in advance by the formulation of a suitable fitting function. We anticipate this could offer great advantage when deployed in large-throughput user facilities.
\section*{References}
\bibliography{iopart-num}

\end{document}